\documentstyle[11pt,amsmath,amsfonts,amssymb,epsfig]{article}
\begin{document}
\title{Semi-inclusive production of pions in DIS \\
 and $\bar d - \bar u$ asymmetry}

   \author{V. Uleshchenko$^{1,2}$, A. Szczurek$^{1}$ and J. Speth$^{3}$  \\
   {\it $^{1}$ Institute of Nuclear Physics, PL-31-342 Cracow, Poland  } \\
   {\it $^{2}$ Institute for Nuclear Research, 03-028 Kiev, Ukraine  } \\
   {\it $^{3}$ Institut f\"ur Kernphysik, KFA, J\"ulich, Germany} \\ }

\maketitle

\begin{abstract}
We discuss the role of some nonpartonic effects which lead to
$N_p^{\pi^+} \ne N_n^{\pi^+}$ and
$N_p^{\pi^-} \ne N_n^{\pi^-}$ and may therefore modify the conclusion
on the $\bar d - \bar u$ asymmetry extracted
from semi-inclusive production of pions in DIS.
Quantitative estimations for resolved photon and exclusive $\rho^0$
are given as examples.
The results are discussed in the context of recent HERMES data.
\end{abstract}

\vspace{0.5cm}


\section{Introduction}

In order to understand the nature
of the Gottfried Sum Rule violation \cite{NMC_GSR} two different
Drell-Yan experiments were performed \cite{NA51,E866}.
The integrated result for
the asymmetry from a more complete Fermilab experiment \cite{E866} is
$\int_0^1 [ \bar d - \bar u ] \; dx$ = 0.09 $\pm$ 0.02, to be
compared with the NMC result:
$\int_0^1 [ \bar d - \bar u ] \; dx$ = 0.148 $\pm$ 0.039.
The NMC integral asymmetry apears slightly bigger. It was suggested
recently \cite{SU_part_viol} that the difference can be
partly due to large higher-twist effects for $F_2^p - F_2^n$.
Recently the HERMES collaboration \cite{HERMES_dbar_ubar} used
semi-inclusive unpolarized production of pions to extract the asymmetry.

We discuss briefly a disturbing role of some nonpartonic
processes which cloud the extraction of the asymmetry from
semi-inclusive DIS.

\section{Quark-parton model approach}

In the quark-parton model (see Fig.1a) the generalized
semi-inclusive structure function can be written as
\begin{equation}
{\cal F}_2^{N \rightarrow \pi}(x,Q^2,z) =
\sum_f e_f^2 x q_f(x,Q^2) \cdot D_{f \rightarrow \pi}(z) \; ,
\label{semi_parton}
\end{equation}
where the sum runs over the quark/antiquark flavours $f=u,d,s$,
$q_f$ are quark distribution functions and $D_{f \rightarrow \pi}(z)$
are so-called fragmentation functions.

The isospin and charge conjugation symmetries allow to reduce
the number of fragmentation functions to two: favoured $D_{+}(z)$
and unfavoured $D_{-}(z)$ \footnote{A third type of fragmentation functions
$D_s(z)$ for strange quarks does not enter the quantity analyzed here
(\ref{exp_extract}).}.

In the QPM one can combine semi-inclusive
cross sections for the production of $\pi^+$ and
$\pi^-$  on proton and neutron targets
to isolate a quantity sensitive to the flavour asymmetry
\cite{HERMES_dbar_ubar}
\begin{equation}
\frac{ \bar d(x) - \bar u(x) }{ u(x) - d(x) } =
\frac{J(z) [1 - r(x,z)] - [1 + r(x,z)]}
     {J(z) [1 - r(x,z)] + [1 + r(x,z)]} \; ,
\label{exp_extract}
\end{equation}
where $J(z) = \frac{3}{5}
\frac{1 + D_{-}(z)/D_{+}(z)}{1 - D_{-}(z)/D_{+}(z)} $
and $r(x,z) = \frac{N_p^{\pi^-}(x,z) - N_n^{\pi^-}(x,z)}
                   {N_p^{\pi^+}(x,z) - N_n^{\pi^+}(x,z)}$.
In the absence of other mechanisms the equation can be used
to extract the $x$-dependence of the difference $\bar d - \bar u$.

In order to demonstrate the effect of nonpartonic
components on the extraction of $\bar d - \bar u$ asymmetry
we need to fix the effective fragmentation functions which the
main partonic term will be calculated with.
Most of the model fragmentation functions were constructed
in the context of $e^+ e^-$ pion production data, where amount
of $\pi^+$ and $\pi^-$ produced is equal, and do not manage
to describe separately multiplicity distributions of positive
and negative pions in DIS \cite{SUS00}.
Besides separate yields of $\pi^+$ and $\pi^-$ the QPM formula
(\ref{exp_extract}) directly depends also on the ratio of
unfavoured and favoured fragmentation functions.

Surprisingly only a rather old Field-Feynman parametrization \cite{FF77}
provides a good representation of the available $e p$ pion production data
in the HERMES kinematical region \cite{SUS00}. This parametrization
will be used in the following analysis.

\section{Nonpartonic components}

For small $Q^2$, as in the case of the HERMES experiment,
some mechanisms of nonpartonic origin may become important too.
For instance the virtual photon can interact with the nucleon
via its intermediate hadronic state. Such a mechanism can be described
within the vector dominance model (VDM). The photon could also fluctuate
into a pair of pions, where both or one of them interact with the nucleon.
Some exclusive processes can produce pions directly or as decay
products of heavier mesons.

\begin{figure}
\begin{center}
\mbox{
\epsfysize 7.0cm
\epsfbox{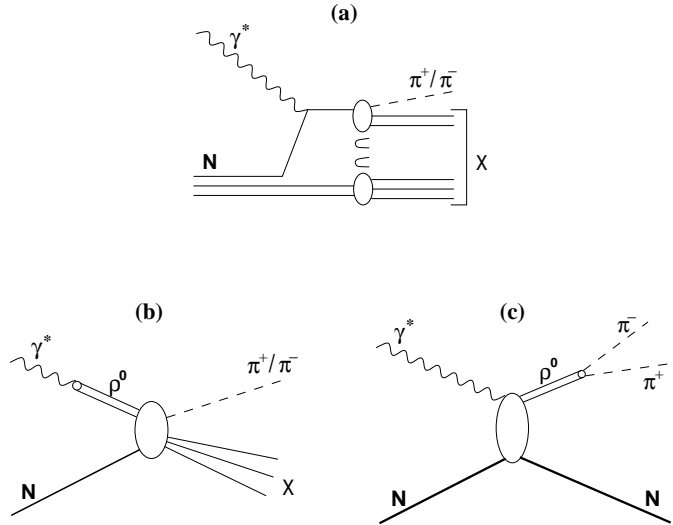}
}
\caption{\it
Different possible mechanisms of the pion production:
(a) quark fragmentation,
(b) VDM contribution,
(c) elastic production of the $\rho^0$ meson and its decay.}
\end{center}
\end{figure}
To our best knowledge none of such processes has been investigated
in the literature. Their influence on the extracted $\bar d - \bar u$
asymmetry also remains unknown. Here for illustration we discuss only
two of them.

\subsection{VDM contribution}

Let us start from the VDM component (see Fig.1b).
It was shown recently that the inclusion of the VDM contribution
and a related modification of the partonic component help
to understand the behaviour of structure functions $F_2^p$ and $F_2^d$
at small $Q^2$ and broad range of Bjorken-$x$ \cite{SU_inclusive}.
This model was confirmed by a recent analysis of
the $Q^2$-dependence of the world data for
$F_2^p - F_2^n$ \cite{SU_part_viol}.
The model for inclusive structure functions
\cite{SU_inclusive} can be generalized
to semi-inclusive production of pions:
\begin{eqnarray}
 {\cal F}_2^{N \rightarrow \pi}(x,Q^2,z) &=&
\frac{Q^2}{Q^2+Q_0^2}
\sum_f e_f^2 x q_f(x,Q^2) \cdot D_{f \rightarrow \pi}^{eff}(z) \nonumber \\
&+& \frac{Q^2}{\pi}
\sum_{V} \frac{1}{\gamma_V^2}
\frac{\sigma_{VN \rightarrow \pi X}(s^{1/2}) M_V^4}{(Q^2+M_V^2)^2}
\Omega_V(x,Q^2).
\label{SU_semi}
\end{eqnarray}
The second sum above runs over vector mesons $V = \rho^0, \omega, \Phi$
and $\Omega_V$ decribes a correction factor due to finite fluctuation time
of virtual photon into vector mesons for large $x$ \cite{SU_inclusive}.

The inclusive cross section for pion production in vector meson
($\rho^0, \omega, \phi$) scattering off proton and neutron is not
known experimentally.
In analogy to the total cross section the pion production inclusive
cross section $\rho^0 N \rightarrow \pi^{\pm} X$ can be estimated as:
\begin{eqnarray}
\sigma(\rho^0 p \rightarrow \pi^{\pm} X) &\approx&
 1/2 \; [ \sigma(\pi^+ p \rightarrow \pi^{\pm} X)
         + \sigma(\pi^- p \rightarrow \pi^{\pm} X) ] \; ,
\nonumber \\
\sigma(\rho^0 n \rightarrow \pi^{\pm} X) &\approx&
  1/2 \; [ \sigma(\pi^+ n \rightarrow \pi^{\pm} X)
         + \sigma(\pi^- n \rightarrow \pi^{\pm} X) ]
\; .
\end{eqnarray}
Experimental data from the ABBCCHW collaboration \cite{Bosetti73} at
$p_{lab}^{\pi}$ = 8, 16 GeV correspond approximately to the range
of the HERMES experiment \cite{HERMES_dbar_ubar}.

However, the experimental spectra for $\pi^{\pm} p \rightarrow \pi^{\pm} X$
contain components due to peripheral processes, which are specific,
different for different beams.
We wish to note that peripheral processes in the
$\pi^+ p \rightarrow \pi^+ X$ and $\pi^- p \rightarrow \pi^- X$
reactions do not contribute to the
$\rho^0 p \rightarrow \pi^{\pm} X$ reaction and should be eliminated;
only nondiffractive components for $\pi p \rightarrow \pi X$
reactions should be taken into account.
This requires a physically motivated parametrization
of the $ \pi + N \rightarrow \pi + X$ data.
Therefore we have parametrized the experimental differential cross sections
for four different reactions
$\pi^{\pm} p \rightarrow \pi^{\pm} X$ from \cite{Bosetti73} as
a sum of central and peripheral components
\begin{equation}
\frac{d \sigma}{dx_F dp_{\perp}^2} =
\frac{d \sigma^{cen}}{dx_F dp_{\perp}^2} +
\frac{d \sigma^{per}}{dx_F dp_{\perp}^2}
\; .
\end{equation}
The details of the analysis will be given elsewhere \cite{SUS00}.
Because the CM-energy of the ABBCCHW collaboration is very similar
to that of the HERMES experiment,
we believe that the parametrization is suitable in the
limiting range
of energy relevant for the HERMES experiment \cite{HERMES_dbar_ubar}.
The cross sections on the neutron can be obtained
from those on the proton by assuming isospin symmetry for the
hadronic reactions.

The analysis of experimental data \cite{Bosetti73}
combined with the assumption of isospin symmetry strongly indicate that
for the {\it nondiffractive} components \cite{SUS00}
\begin{equation}
\sigma(\rho^0 p \rightarrow \pi^{\pm} X) \ne
\sigma(\rho^0 n \rightarrow \pi^{\pm} X) \; .
\end{equation}
This automatically means that the VDM contribution modifies the r.h.s.
of Eq.(\ref{exp_extract}).

\begin{figure}
\begin{center}
\mbox{
\epsfysize 8.0cm
\epsfbox{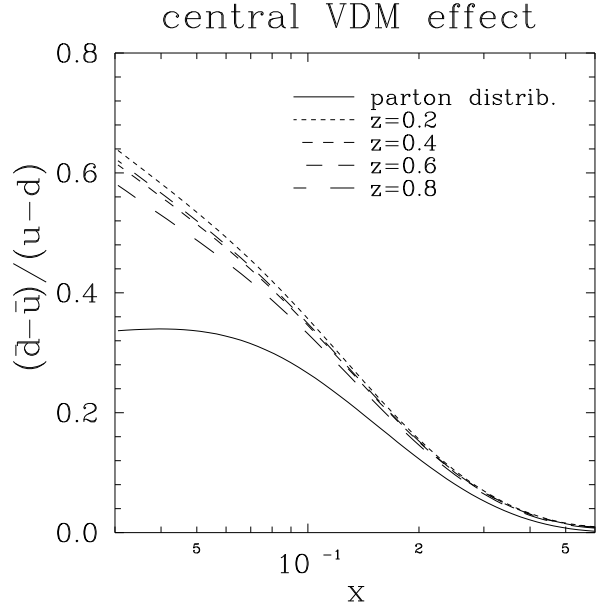}
}
\caption{\it
The "true" (solid) and the modified by the central VDM contribution
$\frac{\bar d - \bar u}{u - d}$ calculated according to l.h.s. and r.h.s.
of Eq.(\ref{exp_extract}), respectively, as a function of Bjorken-$x$
for different  values of $z$ and $W$ = 5 GeV.}
\end{center}
\end{figure}
In Fig.2 we show a modification of the measured HERMES quantity
$\frac{\bar d - \bar u}{u - d}$ due to the VDM component.
In the present calculation the photon-proton CM energy was fixed at
the average HERMES value $W$ = 5.0 GeV$^2$
and the quark fragmentation component was rescalled by a factor
$\frac{Q^2}{Q^2+Q_0^2}$ (see Eq.(\ref{SU_semi})).
The solid line represents $\frac{\bar d - \bar u}{u - d}$
obtained directly from the parton distributions \cite{GRV94}.
As can be seen from the figure the r.h.s. of Eq.(\ref{exp_extract})
clearly deviates from the partonic result.
Thus, if not canceled by other effects, the quark flavour asymmetry
extracted from semi-inclusive experiments in the simple QPM approach
seems to be highly overestimated.

\subsection{Exclusive $\rho$ meson production}

The exclusive meson production
$\gamma^* N \rightarrow M N'$ is not included
in the fragmentation formalism and
may also modify the extraction of $\bar d - \bar u$ asymmetry.
The pion exclusive channels ($M = \pi$) contribute at $ z \approx$ 1,
i.e. outside of the range of the HERMES kinematics
and will be ignored here.
The pions from decays of light vector mesons can be important
in the context of the $\bar d - \bar u$ asymmetry.
The production of $\rho$ mesons ($M = \rho$) seems to be of
particular importance. First of all the $\rho^0 N$ channel is know to
be the dominant exclusive channel in the $\gamma^* N$ scattering.
Secondly, because $\rho^0$ decays predominantly
into two pions it will produce pions with $< z > \sim \frac{1}{2}$.
A detailed calculation \cite{SUS00} shows that the dispersion of
the decay-pion $z$-distribution is large and therefore this effect
could be observed at large $z$ where hadronization rate is rather small.
Below as an example we shall consider the $\rho^0$ elastic production
only.

The elastic $\rho^0$-production contribution
(diagram (c) in Fig.1) to semi-inclusive structure function can
be written formally as
\begin{equation}
{\cal F}_2^{el,\rho^0}(x,Q^2,z) = \frac{Q^2}{4 \pi^2 \alpha}
\cdot \sigma_{\gamma^* N \rightarrow \rho^0 N}(W,Q^2) \cdot
f_{decay}(z)  \; .
\end{equation}

The cross section for proton and neutron target and their difference
can be estimated within the Regge approach as well as
in a QCD inspired quark-exchange model. It is not clear a priori
what is the applicability range of these models.
In this short note we shall try to understand the elastic $\rho^0$
meson production only within the Regge phenomenology.
This requires an analysis of relevant experimental data for
the proton and deuteron targets simultaneously.
While for the proton target there are data in quite a broad kinematical
range of $x$ and $Q^2$ \cite{proton_rho0} (although slightly different
from the HERMES kinematics), there is almost no data for the deuteron target.

We have parametrized the existing experimental
data for exclusive $\rho^0$ production by means of the following simple
Regge-inspired reaction amplitude
\begin{eqnarray}
A_{ \lambda_{N'} \leftarrow \lambda_{N} }
 ^{ \lambda_{V} \leftarrow \lambda_{\gamma} }
(\gamma^* N \rightarrow \rho^0 N; t) &=&
(
i \cdot C_{IP}(t) \left( \frac{s}{s_0} \right)^{\varepsilon} +
\left[ \frac{-1+i}{\sqrt{2}} \right]
\cdot C_{IS}(t) \left( \frac{s}{s_0} \right)^{-1/2}
\nonumber \\
&\pm&
\left[ \frac{-1+i}{\sqrt{2}} \right]
\cdot C_{IV}(t) \left( \frac{s}{s_0} \right)^{-1/2}
)
\nonumber \\
&&\cdot \left[ \frac{m_{\rho}^2}{m_{\rho}^2+Q^2} \right]
\delta_{\lambda_{N'} \lambda_{N}} \delta_{\lambda_{V} \lambda_{\gamma}}
\label{Regge_amplitude}
\end{eqnarray}
with "$+$" for proton and "$-$" for neutron, respectively.
In practical application we shall assume the same t-dependence of
$C_{IP}$, $C_{IS}$ and $C_{IV}$ and take $\Lambda = m_{\rho}$.

The free parameters in Eq.(\ref{Regge_amplitude})
i.e. $\epsilon$, $C_{IP}$ and $C_{IS}+C_{IV}$  have been fitted
to the existing experimental data for $t$-integrated differential
cross section for $\rho^0$ production on hydrogen \cite{proton_rho0}.
In this fit we have limited to the experimental data with
$W >$ 3 GeV (to avoid resonances) and $Q^2 <$ 10 GeV$^2$
(above genuine hard QCD processes should reveal) and
fixed the slope parameter $B$ = 6 GeV$^{-2}$ in exponential
$t$-distribution which is known experimentally.
All other details of the analysis will be discussed in \cite{SUS00}.

From the fit to the proton data we have obtained only a sum
of the coefficients $C_{IS}+C_{IV}$.
The separation into the isoscalar and isovector
contributions cannot be done in a model independent way.
The size of the isovector $a_2$-exchange contribution was estimated
long ago for total photoproduction cross section (see for instance
\cite{Leith78}).
Our Regge model can be applied to both real and virtual photoproduction
and exclusive as well as inclusive case \cite{SUS00}. We have used
the data for real photoproduction total cross section and the data
for exclusive $\omega$ photoproduction to estimate the size of the
isovector amplitude \cite{SUS00} i.e. the strength of
$a_2$-reggeon exchange.

\begin{figure}
\begin{center}
\mbox{
\epsfysize 7.0cm
\epsfbox{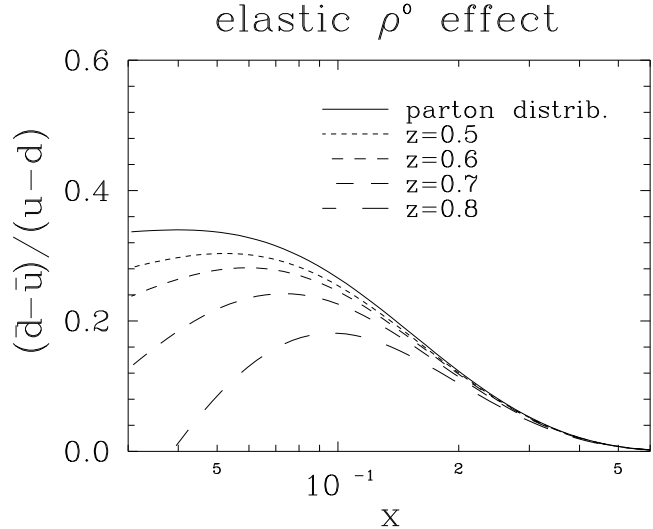}
}
\caption{\it
The "true" (solid) and the modified by the exclusive $\rho^0$
production $(\bar d - \bar u) / (u - d)$ as a function of
Bjorken-$x$ for different $z$ and $W$ = 5 GeV.
}
\end{center}
\end{figure}
Although the difference of the cross sections for exclusive $\rho^0$
photoproduction on the neutron/proton targets calculated with
resulting amplitudes is small, its effect
on the $\frac{\bar d - \bar u}{u-d}$ ratio extracted by the HERMES
collaboration, shown in Fig.3, is not negligible at all.
We show in the figure the "measured" HERMES quantity as a function
of Bjorken-$x$ for a few values of $z$ for a fixed $W$ = 5 GeV.
As in the case of the VDM contribution the QPM term was modified
by the factor $\frac{Q^2}{Q^2+Q_0^2}$ according
to our prescription for inclusive structure functions \cite{SU_inclusive}.

\section{Conclusions}

The semi-inclusive production of pions was recently used to
determine the $\bar u - \bar d$ asymmetry in the nucleon sea.
In the present analysis we have investigated a few
effects beyond the quark-parton model which may influence
the so-extracted asymmetry.

According to our estimation
the interaction of the resolved hadron-like component of the photon
with the nucleon leads to an artificial enhancement of
the measured $\bar d - \bar u$ asymmetry in the region of small
$x$. This enhancement depends on $z$ very weakly.

The elastic production of $\rho^0$ meson is equally important.
This effect, however, modifies the measured $\bar d - \bar u$
asymmetry in the opposite direction from the resolved photon component,
but the two effects cancel only within a narrow interval of $z$.
The elastic $\rho^0$ production makes the r.h.s. of
Eq.(\ref{exp_extract}) $z$-dependent invalidating somewhat averaging
done in \cite{HERMES_dbar_ubar}.

Here we have only shortly discussed two effects. A more detailed
analysis of these two and other effects will be presented elsewhere
\cite{SUS00}.
Finally we wish to conclude that nonperturbative effects beyond QPM
may substantially disturb the extraction of the $\bar d - \bar u$
asymmetry from semi-inclusive production of pions in DIS.
Such an extraction requires a carefull combined analysis including
many nonpartonic effects together with main partonic term.

In addition we would like to point out that some of the effects discussed
in the present paper may also influence the extraction of the polarized
quark distributions from semi-inclusive production of pions in DIS.

\vskip 1cm

{\bf Acknowledgments}
We are indebted to the members of the HERMES collaboration for
discussions of their recent results.
This work was partially supported by the German-Polish
DLR exchange program, grant number POL-028-98.

\end{document}